\documentclass[12pt,a4paper]{article}

\usepackage{afterpage}
\usepackage{amsmath}
\usepackage{amssymb}
\usepackage{color}
\usepackage{graphicx}
\usepackage{hyperref}
\usepackage{authblk}

\title{Versatile linkage: a family of space-conserving strategies for agglomerative hierarchical clustering}

\author[1]{Alberto Fern{\'a}ndez}
\author[2]{Sergio G{\'o}mez}

\affil[1]{Departament d'Enginyeria Qu\'{\i}mica, Universitat Rovira i Virgili, 43007 Tarragona, Spain.
  \href{mailto:alberto.fernandez@urv.cat}{alberto.fernandez@urv.cat}}

\affil[2]{Departament d'Enginyeria Inform\`{a}tica i Matem\`{a}tiques, Universitat Rovira i Virgili, 43007 Tarragona, Spain. \href{mailto:sergio.gomez@urv.cat}{sergio.gomez@urv.cat}}

\date{}

\begin{document}

\maketitle

\begin{abstract}
\footnotesize
\textbf{Abstract:} Agglomerative hierarchical clustering can be implemented with several strategies that differ in the way elements of a collection are grouped together to build a hierarchy of clusters. Here we introduce versatile linkage, a new infinite system of agglomerative hierarchical clustering strategies based on generalized means, which go from single linkage to complete linkage, passing through arithmetic average linkage and other clustering methods yet unexplored such as geometric linkage and harmonic linkage. We compare the different clustering strategies in terms of cophenetic correlation, mean absolute error, and also tree balance and space distortion, two new measures proposed to describe hierarchical trees. Unlike the $\beta$-flexible clustering system, we show that the versatile linkage family is space-conserving.
\end{abstract}


\newpage

\section{Introduction}
\label{sec:introduction}

Agglomerative hierarchical clustering constitutes one of the most widely used methods for cluster analysis. Starting with a matrix of dissimilarities between a set of elements, each element is first assigned to its own cluster, and the algorithm sequentially merges the more similar clusters until a complete hierarchy of clusters is obtained \cite{Sneath1973, Gordon1999}. This method requires the definition of the dissimilarity (or distance) between clusters, using only the original distances between their constituent elements. The way these distances are defined leads to distinct strategies of agglomerative hierarchical clustering. To name just two such clustering strategies, single linkage usually leads to an elongate growth of clusters, while complete linkage generally leads to tight clusters that join others with difficulty. Average linkage clustering strategies were developed by Sokal and Michener to avoid the extreme cases produced by single linkage and complete linkage \cite{Sokal1958}. They require the calculation of some kind of average distance between clusters; average linkage, for instance, calculates the arithmetic average of all the distances between members of the clusters.

More than fifty years ago, Lance and Williams introduced a formula for integrating several agglomerative hierarchical clustering strategies into a single system \cite{Lance1966}. Based on this formula they proposed $\beta$-flexible clustering \cite{Lance1967}, a generalized clustering procedure that provides an infinite number of hierarchical clustering strategies just varying a parameter $\beta$. Similarly, in this work we introduce \textit{versatile linkage}, a new parameterized family of agglomerative hierarchical clustering strategies that go from single linkage to complete linkage, passing through arithmetic average linkage and other clustering strategies yet unexplored such as geometric linkage and harmonic linkage.

Both $\beta$-flexible clustering and versatile linkage are presented here using variable-group methods \cite{Sokal1958, Fernandez2008} that, unlike pair-group methods, admit any number of new members simultaneously into groups. In the case of pair-group methods the resulting hierarchical tree is called a dendrogram, which is built upon bifurcations, while in the case of variable-group methods the resulting hierarchical tree is called a \textit{multidendrogram} \cite{Fernandez2008}, which consists of multifurcations, not necessarily binary ones. Here we use the variable-group algorithm introduced in \cite{Fernandez2008} that solves the non-uniqueness problem, also called the ties in proximity problem, found in pair-group algorithms \cite{Sneath1973, Hart1983, Day1984}. This problem arises when there are more than two clusters separated by the same minimum distance during the agglomerative process. Pair-group algorithms break ties between distances choosing a pair of clusters, usually at random. However, different output dendrograms are possible depending on the criterion used to break ties. Moreover, very frequently results depend on the order of the elements in the input data file, what is an undesired effect in hierarchical clustering except for the case of contiguity-constrained hierarchical clustering, which is used to obtain a hierarchical clustering that takes into account the ordering on the input elements. The variable-group algorithm used here always gives a uniquely determined solution grouping more than two clusters at the same time when ties occur, and when there are no ties it gives the same results as the pair-group algorithm.

Section~\ref{sec:flexible} reviews the $\beta$-flexible family of hierarchical clustering strategies, while Section~\ref{sec:versatile} introduces the versatile linkage family. Four case studies are used in Section~\ref{sec:analysis} to perform a descriptive analysis of different hierarchical clustering strategies in terms of cophenetic correlation, mean absolute error, and the proposed new measures of space distortion and tree balance. Finally, some concluding remarks are given in Section~\ref{sec:conclusions}.

\section{$\beta$-Flexible Clustering}
\label{sec:flexible}

In any procedure implementing an agglomerative hierarchical clustering strategy, given a set of individuals $\Omega = \{x_{1},x_{2},\ldots,x_{n}\}$, initially each individual forms a singleton cluster, $\{x_{i}\}$, and the distances $D(\{x_{i}\},\{x_{j}\})$ between singleton clusters are equal to the dissimilarities between individuals, $d(x_{i},x_{j})$. During the subsequent iterations of the procedure, the distances $D(X_{I},X_{J})$ are computed between any two clusters $X_{I}=\bigcup_{i \in I}X_{i}$ and $X_{J}=\bigcup_{j \in J}X_{j}$, each one of them made up of several subclusters $X_{i}$ and $X_{j}$ indexed by $I=\{i_{1},i_{2},\ldots,i_{p}\}$ and $J=\{j_{1},j_{2},\ldots,j_{q}\}$, respectively. Lance and Williams introduced a formula for integrating several agglomerative hierarchical clustering strategies into a single system \cite{Lance1966}. The variable-group generalization of Lance and Williams' formula, compatible with the fusion of more than two clusters simultaneously, is:
\begin{eqnarray}
  \label{eq:Lance_Williams}
  \lefteqn{ D(X_{I},X_{J}) = \sum_{i \in I} \sum_{j \in J} \alpha_{ij} D(X_{i},X_{j}) + \mbox{}} \nonumber \\
    & & + \sum_{i \in I} \sum_{\substack{i' \in I \\ i'>i}} \beta_{ii'} D(X_{i},X_{i'})
        + \sum_{j \in J} \sum_{\substack{j' \in J \\ j'>j}} \beta_{jj'} D(X_{j},X_{j'})\,,
\end{eqnarray}
where the values of the parameters $\alpha_{ij}$, $\beta_{ii'}$ and $\beta_{jj'}$ determine the nature of the clustering strategy \cite{Fernandez2008}. This formula is combinatorial \cite{Lance1967}, i.e., the distance $D(X_{I},X_{J})$ can be calculated from the distances $D(X_{i},X_{j})$, $D(X_{i},X_{i'})$ and $D(X_{j},X_{j'})$ obtained from the previous iteration and it is not necessary to keep the initial distance matrix $d(x_{i},x_{j})$ during the whole clustering process.

Based on Equation~\ref{eq:Lance_Williams}, Lance and Williams \cite{Lance1967} proposed an infinite system of agglomerative hierarchical clustering strategies defined by the constraint
\begin{equation}
  \underbrace{\sum_{i \in I} \sum_{j \in J} \alpha_{ij}}_{\alpha}
  + \underbrace{\sum_{i \in I} \sum_{\substack{i' \in I \\ i'>i}} \beta_{ii'}
              + \sum_{j \in J} \sum_{\substack{j' \in J \\ j'>j}} \beta_{jj'}}_{\beta} = 1\,,
\end{equation}
where $-1 \leqslant \beta \leqslant +1$ generates a whole system of hierarchical clustering strategies for the infinite possible values of $\beta$. Given a value of $\beta$, the value for $\alpha_{ij}$ can be assigned following a weighted approach as in the original $\beta$-flexible clustering based on WPGMA (weighted pair-group method using arithmetic mean) and introduced by Lance and Williams \cite{Lance1966}, or it can be assigned following an unweighted approach as in the $\beta$-flexible clustering based on UPGMA (unweighted pair-group method using arithmetic mean) and introduced by Belbin \textit{et al.} \cite{Belbin1992}. The standard WPGMA and UPGMA strategies are obtained from weighted and unweighted $\beta$-flexible clustering, respectively, when $\beta$ is set equal to $0$. The difference between weighted and unweighted methods lies in the weights assigned to individuals and clusters during the agglomerative process: weighted methods assign equal weights to clusters, while unweighted methods assign equal weights to individuals. In unweighted $\beta$-flexible clustering the value for $\alpha_{ij}$ is determined proportionally to $|X_{i}||X_{j}|$:
\begin{equation}
  \alpha_{ij} = \frac{|X_{i}||X_{j}|}{|X_{I}||X_{J}|} (1 - \beta)\,,
\end{equation}
where $|X_{i}|$ and $|X_{j}|$ are the number of individuals in subclusters $X_{i}$ and $X_{j}$, respectively, and $|X_{I}|$ and $|X_{J}|$ are the number of individuals in clusters $X_{I}$ and $X_{J}$, i.e., $|X_{I}| = \sum_{i \in I} |X_{i}|$ and $|X_{J}| = \sum_{j \in J} |X_{j}|$. In a similar way, the value for $\beta_{ii'}$ is calculated proportionally to $|X_{i}||X_{i'}|$, and the value for $\beta_{jj'}$ proportionally to $|X_{j}||X_{j'}|$:
\begin{eqnarray}
  \beta_{ii'} & = & \frac{|X_{i}||X_{i'}|}{\sigma_{I} + \sigma_{J}} \beta\,, \\
  \sigma_{I} & = & \sum_{i \in I} \sum_{\substack{i' \in I \\ i'>i}} |X_{i}||X_{i'}| = \frac{1}{2} \left( |X_{I}|^{2} - \sum_{i \in I} |X_{i}|^{2} \right)\,.
\end{eqnarray}
The corresponding values for weighted $\beta$-flexible clustering are:
\begin{eqnarray}
  \alpha_{ij} & = & \frac{1}{|I||J|} (1 - \beta)\,, \\
  \beta_{ii'} & = & \frac{1}{\sigma_{I} + \sigma_{J}} \beta\,, \\
  \sigma_{I} & = & \frac{|I| \left( |I| - 1 \right)}{2} = \frac{|I|^{2} - |I|}{2}\,,
\end{eqnarray}
where $|I|$ and $|J|$ are the number of subclusters contained in clusters $X_{I}$ and $X_{J}$, respectively. These formulas derive from the unweighted ones when we take $|X_{i}| = 1$, $\forall i \in I$, and $|X_{j}| = 1$, $\forall j \in J$.

\section{Versatile Linkage}
\label{sec:versatile}

Arithmetic average linkage clustering iteratively forms clusters made up of previously formed subclusters, based on the arithmetic mean distances between their member individuals; for simplicity and to avoid confusion, we will denote it \textit{arithmetic linkage} instead of the standard term \textit{average linkage}. Substituting the arithmetic means by generalized means, also known as power means, this clustering strategy can be extended to any finite power $p \neq 0$:
\begin{eqnarray}
  \lefteqn{ D_{p}(X_{I},X_{J}) = \left( \frac{1}{|X_{I}||X_{J}|} \sum_{x \in X_{I}} \sum_{y \in X_{J}} [d(x,y)]^{p} \right)^{1/p} = } \nonumber \\
    & & = \left( \frac{1}{|X_{I}||X_{J}|} \sum_{i \in I} \sum_{j \in J} |X_{i}||X_{j}| [D_{p}(X_{i},X_{j})]^{p} \right)^{1/p}\,.
  \label{eq:unweighted_versatile}
\end{eqnarray}
We call this new system of agglomerative hierarchical clustering strategies as \textit{versatile linkage}. As in the case of $\beta$-flexible clustering, versatile linkage provides a way of obtaining an infinite number of clustering strategies from a single formula. The second equality in Equation~\ref{eq:unweighted_versatile} shows that versatile linkage can be calculated using a combinatorial formula, from the distances $D_{p}(X_{i},X_{j})$ obtained during the previous iteration, in the same way as Lance and Williams' recurrence formula given in Equation~\ref{eq:Lance_Williams}.

The decision of what power~$p$ to use could be taken in agreement with the type of distance employed to measure the initial dissimilarities between individuals. For instance, if the initial dissimilarities were calculated using a generalized distance of order~$p$, then the natural agglomerative clustering strategy would be versatile linkage with the same power~$p$. However, this procedure does not guarantee that the dendrogram obtained is the best according to other criteria, e.g., cophenetic correlation, mean absolute error, space distortion or tree balance, see Section~\ref{sec:analysis}. A better approach consists in scanning the whole range of parameters~$p$, calculate the preferred descriptors of the corresponding dendrograms, and decide if it is better to substitute the natural parameter~$p$ by another one. This is especially important when only the dissimilarities between individuals are available, without coordinates for the individuals, as is common in multidimensional scaling problems, or when the dissimilarities have not been calculated using generalized means.

\subsection{Particular Cases}

The generalized mean contains several well-known particular cases, depending on the value of the power $p$, that deserve special attention. Some of them reduce versatile linkage to the most commonly used methods, while others emerge naturally as deserving further attention:
\begin{itemize}
  \item In the limit when $p \rightarrow -\infty$, versatile linkage becomes single linkage (SL):
\begin{equation}
  D_{\min}(X_{I},X_{J}) = \min_{x \in X_{I}} \, \min_{y \in X_{J}} \, d(x,y) = \min_{i \in I} \, \min_{j \in J} \, D_{\min}(X_{i},X_{j})\,.
\end{equation}
  \item In the limit when $p \rightarrow +\infty$, versatile linkage becomes complete linkage (CL):
\begin{equation}
  D_{\max}(X_{I},X_{J}) = \max_{x \in X_{I}} \, \max_{y \in X_{J}} \, d(x,y) = \max_{i \in I} \, \max_{j \in J} \, D_{\max}(X_{i},X_{j})\,.
\end{equation}
\end{itemize}
There are also three other particular cases that can be grouped together as \textit{Pythagorean linkages}:
\begin{itemize}
  \item When $p = +1$, the generalized mean is equal to the arithmetic mean and \textit{arithmetic linkage} (AL), i.e.\ the standard average linkage or UPGMA, is recovered.
  \item When $p = -1$, the generalized mean is equal to the harmonic mean and, therefore, \textit{harmonic linkage} (HL) is obtained.
  \item In the limit when $p \rightarrow 0$, the generalized mean tends to the geometric mean. Hence, the distance definition for \textit{geometric linkage} (GL) is:
\begin{eqnarray}
  \lefteqn{ D_{\mathrm{geo}}(X_{I},X_{J}) = \left( \prod_{x \in X_{I}} \prod_{y \in X_{J}} d(x,y) \right)^{1/(|X_{I}||X_{J}|)} = } \nonumber \\
    & & = \left( \prod_{i \in I} \prod_{j \in J} [D_{\mathrm{geo}}(X_{i},X_{j})]^{|X_{i}||X_{j}|} \right)^{1/(|X_{I}||X_{J}|)}\,.
\end{eqnarray}
\end{itemize}

\begin{table}[tb!]
  \begin{center}
  \caption{Sample pairwise distances between four individuals.}
  \begin{tabular}{|l|rrrr|}
    \cline{2-5}
    \multicolumn{1}{c|}{} & Alice & Bob & Carol & Dave \\
    \hline
    Alice &  0 &  7 & 16 & 28 \\
    Bob   &    &  0 &  9 & 21 \\
    Carol &    &    &  0 & 12 \\
    Dave  &    &    &    &  0 \\
    \hline
  \end{tabular}
  \label{tab:toy_dataset}
  \end{center}
\end{table}

To show the effects of varying the power $p$ in versatile linkage clustering, we have built a small dataset with four individuals: Alice, Bob, Carol and Dave, which lay on a straight line, separated between them by distances equal to $7$, $9$ and $12$ units, respectively. Table~\ref{tab:toy_dataset} gives the pairwise distances between the four individuals, and Figure~\ref{fig:versatile} shows some multidendrograms obtained varying the power $p$ in versatile linkage clustering. Alice and Bob are always grouped together forming the first binary cluster, at a distance equal to $7.00$. For values of the exponent $p \in (-\infty,0)$, the Alice-Bob cluster is joined with Carol's singleton cluster at distances that range between $9.00$ and $12.00$. More precisely, this distance takes values $9.00$ for SL ($p \rightarrow -\infty$), $11.52$ for HL ($p = -1$) and $12.00$ when we approach GL ($p \rightarrow 0^{-}$). For larger values of the exponent, $p > 0$, this distance becomes larger than $12.00$, thus Carol joins instead in a cluster with Dave at their distance $12.00$. The remaining cluster for $p \in (-\infty,0)$, which joins the Alice-Bob-Carol cluster with Dave, happens at heights $12.00$ (SL), $18.00$ (HL) and $19.18$ ($p \rightarrow 0^{-}$), respectively. For the range $p \in (0,+\infty)$, the clusters Alice-Bob and Carol-Dave join at heights $17.06$ ($p \rightarrow 0^{+}$), $18.50$ (AL) and $28.00$ (CL), respectively.

\afterpage{
\begin{figure}[tb!]
  \begin{center}
    \begin{tabular}{cccccc}
      \includegraphics[width=0.14\textwidth]{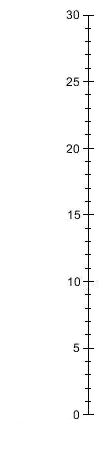} &
      \includegraphics[width=0.14\textwidth]{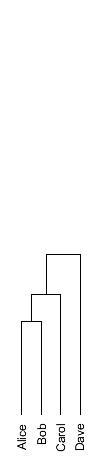} &
      \includegraphics[width=0.14\textwidth]{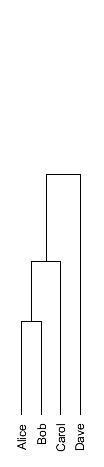} &
      \includegraphics[width=0.14\textwidth]{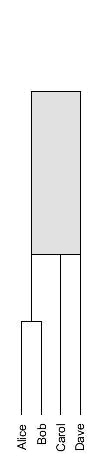} &
      \includegraphics[width=0.14\textwidth]{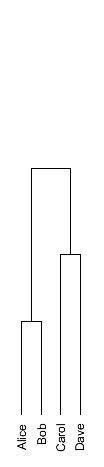} &
      \includegraphics[width=0.14\textwidth]{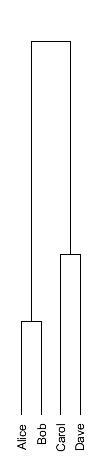} \\
      &
      $\scriptstyle p \rightarrow -\infty$ {\scriptsize (SL)} &
      $\scriptstyle p = -1$ {\scriptsize (HL)} &
      $\scriptstyle p = 0$ {\scriptsize (GL)} &
      $\scriptstyle p = +1$ {\scriptsize (AL)} &
      $\scriptstyle p \rightarrow +\infty$ {\scriptsize (CL)}
    \end{tabular}
	\caption{Effects of varying the power $p$ in versatile linkage clustering for the sample distances in Table~\ref{tab:toy_dataset}. Computations performed using the MultiDendrograms$^{3,4}$ software \cite{Gomez2018}, with the precision parameter equal to 2 significant decimal digits. When $p = 0$ (GL), the gray band shows the existence of a tie between distances.}
	\label{fig:versatile}
  \end{center}
\end{figure}
\setcounter{footnote}{3}
\footnotetext{MultiDendrograms: \protect\url{http://deim.urv.cat/~sergio.gomez/multidendrograms.php}}
\setcounter{footnote}{4}
\footnotetext{In MultiDendrograms, to avoid the infinite range of the exponent $p$, a sigmoidal transformation is performed such that the parameter used is within the range $[-1.0,+1.0]$, with values $-1.0$, $-0.1$, $0.0$, $+0.1$ and $+1.0$ representing SL, HL, GL, AL and CL, respectively.}
}

GL ($p = 0$) lays between these two structurally different dendrograms, represented as ``(((Alice,Bob),Carol),Dave)'' and ``((Alice,Bob),(Carol,Dave))''. Using pair-group agglomerative clustering methods, we would assign one of these two possible dendrograms to GL, thus breaking the tied pairs (Alice,Bob)-Carol and Carol-Dave (both at distance $12.00$) randomly; this is an example of the ties in proximity (non-uniqueness) problem mentioned above. With the variable-group approach \cite{Fernandez2008}, we join them at once forming the multidendrogram ``((Alice,Bob),Carol,Dave)'', where the three clusters join at distance $12.00$, with a band going up to distance $24.25$ to represent the heterogeneity of the new cluster, $24.25$ being the distance between the clusters (Alice,Bob) and Dave (see middle multidendrogram in Figure~\ref{fig:versatile}). This simple example shows the ability of versatile linkage to cover structurally different hierarchical clustering structures, including at the same time the traditionally important methods of SL, AL and CL.

\subsection{Weighted Versatile Linkage}

Weighted clustering was introduced by Sokal and Michener \cite{Sokal1958} in an attempt to give merging branches in a hierarchical tree equal weight regardless of the number of individuals carried on each branch. Such a procedure weights the individuals unequally, contrasting with unweighted clustering that gives equal weight to each individual in the clusters.

In weighted versatile linkage strategies, the distance between two clusters $X_{I}$ and $X_{J}$ is calculated by taking the generalized mean of the pairwise distances, not between individuals in the initial distance matrix, but between component subclusters in the matrix used during the previous iteration of the procedure, thus Equation~\ref{eq:unweighted_versatile} being replaced by:
\begin{equation}
  \label{eq:weighted_versatile}
  D_{p}(X_{I},X_{J}) = \left( \frac{1}{|I||J|} \sum_{i \in I} \sum_{j \in J} \left[ D_{p}(X_{i},X_{j}) \right]^{p} \right)^{1/p}\,.
\end{equation}

\subsection{Absence of inversions}

Versatile linkage strategies are monotonic, that is, they do not produce inversions. An inversion or reversal appears in a hierarchy when the hierarchy contains two clusters $X$ and $Y$ for which $X \subset Y$ but the height of cluster $X$ is higher than the height of cluster $Y$ \cite{Murtagh1985, Morgan1995}. Inversions make hierarchies difficult to interpret, specially if they occur during the last stages of the agglomeration process.

The monotonicity of versatile linkage strategies is explained by the Pythagorean means inequality,
\begin{equation}
  \min \leqslant \mbox{HM} \leqslant \mbox{GM} \leqslant \mbox{AM} \leqslant \max\,,
\end{equation}
where HM stands for the harmonic mean, GM for the geometric mean, and AM for the arithmetic mean. In the general case given by Equations~\ref{eq:unweighted_versatile} and~\ref{eq:weighted_versatile}, the generalized mean inequality holds:
\begin{equation}
  \label{eq:generalized_inequality}
  D_{p}(X_{I},X_{J}) \leqslant D_{q}(X_{I},X_{J})\,, \qquad \forall p < q\,,
\end{equation}
and $D_{p}(X_{I},X_{J}) = D_{q}(X_{I},X_{J})$ if, and only if, the initial distances $d(x,y)$ are equal $\forall x \in X_{I}$ and $\forall y \in X_{J}$. Supposing that at a certain step of the clustering procedure the minimum distance between any two subclusters still to be merged is equal to $\delta$, then the distance $D(X_{i},X_{j})$ between any two subclusters to be included in different clusters, $X_{i} \subseteq X_{I}$ and $X_{j} \subseteq X_{J}$, will be necessarily greater than $\delta$, otherwise subclusters $X_{i}$ and $X_{j}$ would be merged into the same cluster. In particular, $D_{\min}(X_{I},X_{J}) > \delta$. Therefore, taking into account the generalized mean inequality in Equation~\ref{eq:generalized_inequality}, and given that in the limit when $p \rightarrow -\infty$ we have $D_{p}(X_{I},X_{J}) = D_{\min}(X_{I},X_{J})$, we can conclude that $D_{p}(X_{I},X_{J}) > \delta$, $\forall p$, which proves the absence of inversions of versatile linkage strategies.

\section{Descriptive Analysis of Hierarchical Trees}
\label{sec:analysis}

We have selected four case studies, drawn from the UCI Machine Learning Repository \cite{Lichman2013}, for a descriptive analysis of several agglomerative hierarchical clustering strategies. Table~\ref{tab:datasets_characteristics} summarizes the main characteristics of these datasets. The values of the variables in these datasets show different orders of magnitude; therefore, all the variables have been scaled first, and then the corresponding dissimilarity matrices have been built using the Euclidean distance between all pairs of individuals.
\begin{table}[tb!]
  \begin{center}
  \caption{Characteristics of the selected datasets.}
  \begin{tabular}{|lrr|}
    \hline
    Dataset & Instances & Features \\
    \hline \hline
    Breast tissue \cite{Jossinet1996} & 106 & 9 \\
    Iris \cite{Fisher1936}            & 150 & 4 \\
    Wine \cite{Aeberhard1992}         & 178 & 13 \\
    Parkinsons \cite{Little2009}      & 195 & 22 \\
    \hline
  \end{tabular}
  \label{tab:datasets_characteristics}
  \end{center}
\end{table}

For the comparison of the hierarchical clustering strategies, we have chosen the following methods:
$\beta$-flexible with $\beta = +0.9$, to avoid the completely flat hierarchical trees obtained with $\beta = +1$;
versatile linkage with $p \rightarrow -\infty$, i.e., SL;
centroid method;
versatile linkage with $p = -1$, i.e., HL;
versatile linkage with $p \rightarrow 0$, i.e., GL;
versatile linkage with $p = +1$, which is the same as $\beta$-flexible with $\beta = 0$, i.e., AL;
versatile linkage with $p \rightarrow +\infty$, i.e., CL;
Ward's minimum variance method \cite{Ward1963};
and $\beta$-flexible with $\beta = -1$.
This selection includes five variants of versatile linkage, three of them equivalent to traditional methods (SL, AL and CL) and the other two introduced in this work (HL and GL), and three variants of $\beta$-flexible clustering, one of them equivalent to AL.

Weighted and unweighted versions of the hierarchical clustering strategies have been used. Although weighting has no effect on SL and CL, we have included both of them for visual convenience in all the figures depicted next. The software used to run these experiments is MultiDendrograms \cite{Gomez2018}, which from version $5.0$ implements all the hierarchical clustering strategies analyzed here and it also computes the necessary descriptive measures.

\subsection{Cophenetic Correlation}

The cophenetic correlation coefficient (CCC) measures the similarity between the distances in the initial matrix and the distances in the final ultrametric matrix obtained as result of a hierarchical clustering procedure \cite{Sokal1962}. The ultrametric distance between two individuals is represented in a dendrogram by the height at which those two individuals are first joined. The CCC is calculated as the Pearson correlation coefficient between both matrices of distances; thus, the closer to~$1$, the largest their similarity.

\begin{figure}[tb!]
  \begin{center}
    \begin{tabular}{cc}
      \includegraphics[width=0.45\textwidth]{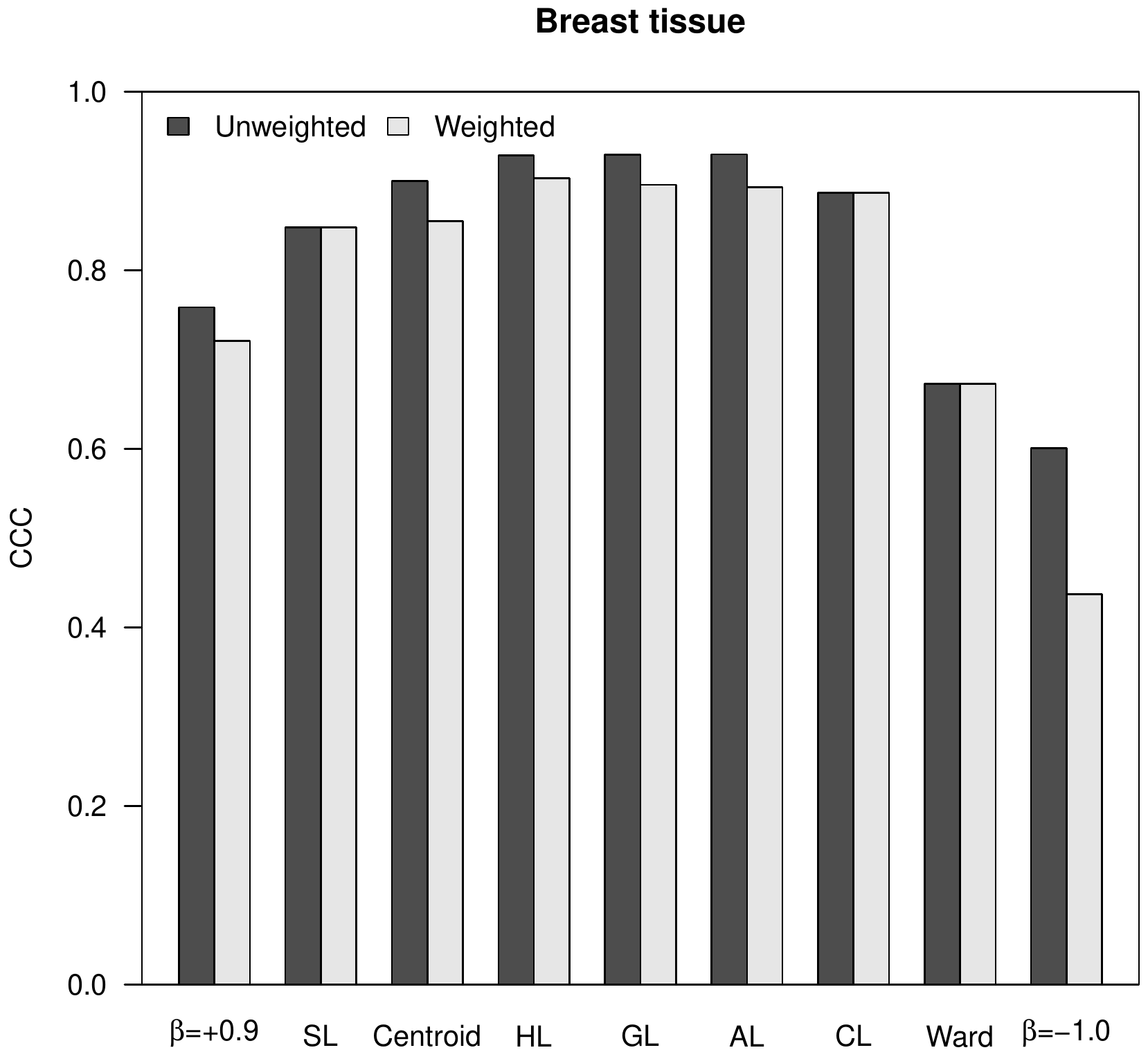} &
      \includegraphics[width=0.45\textwidth]{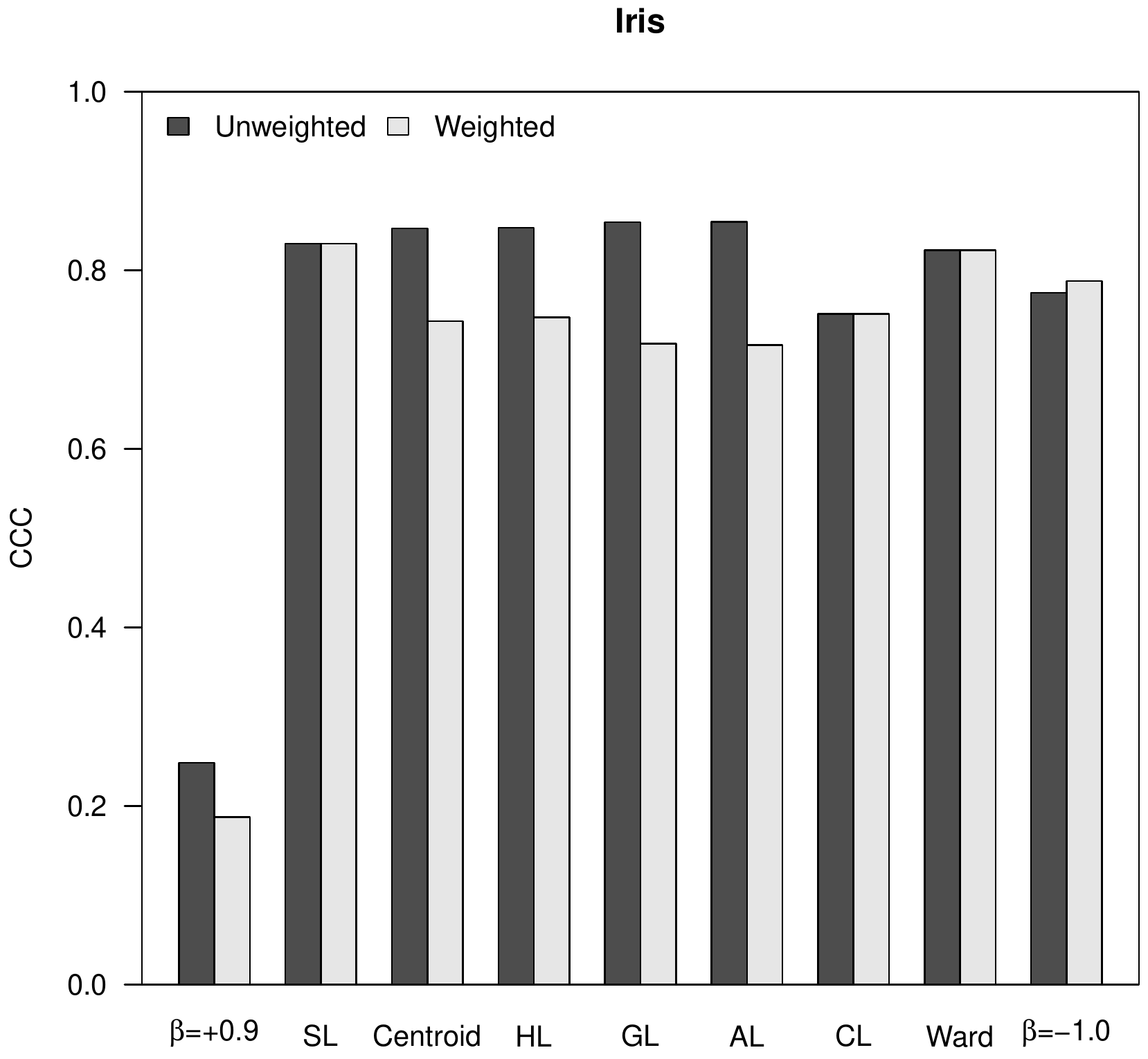} \\
      \includegraphics[width=0.45\textwidth]{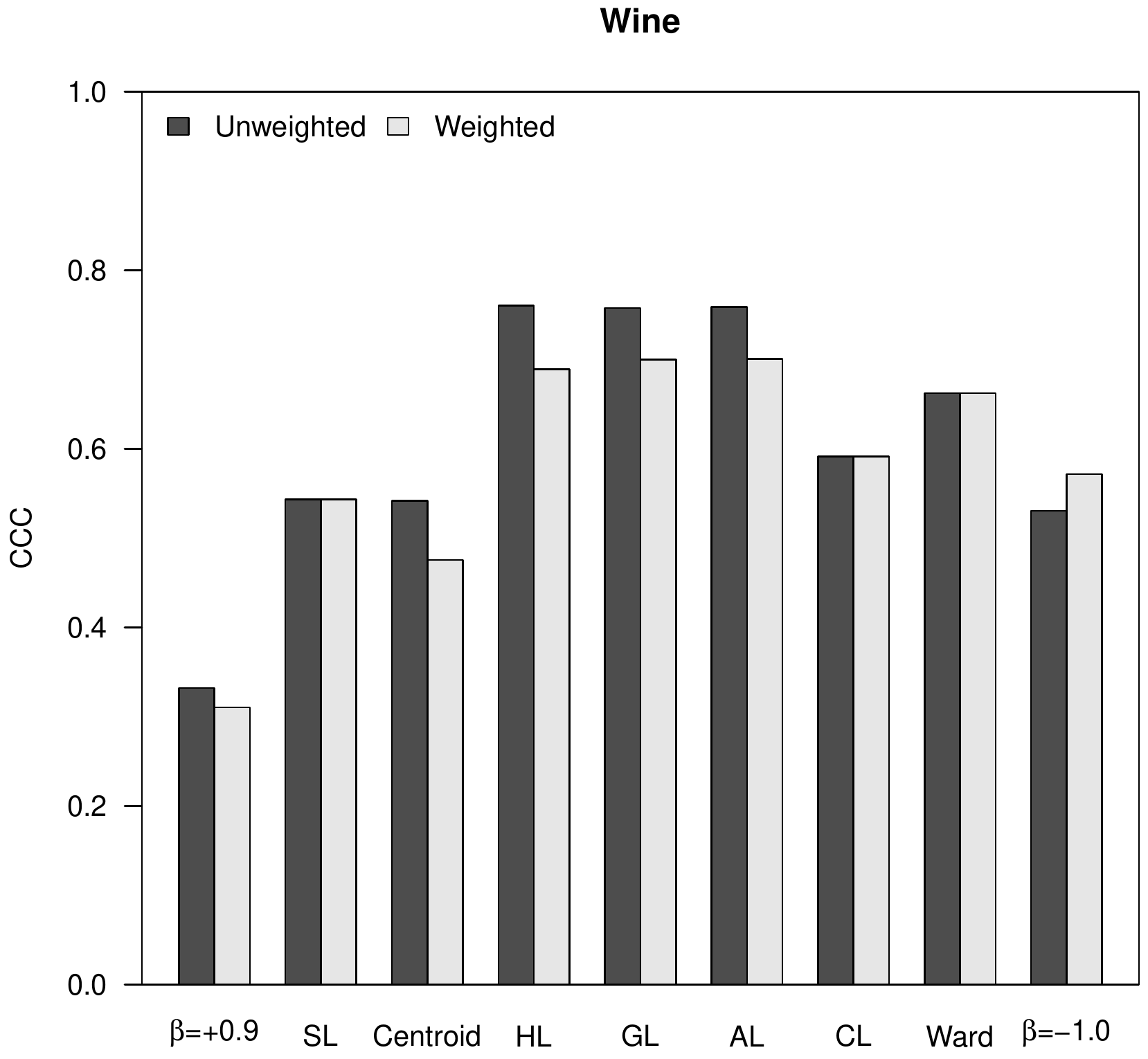} &
      \includegraphics[width=0.45\textwidth]{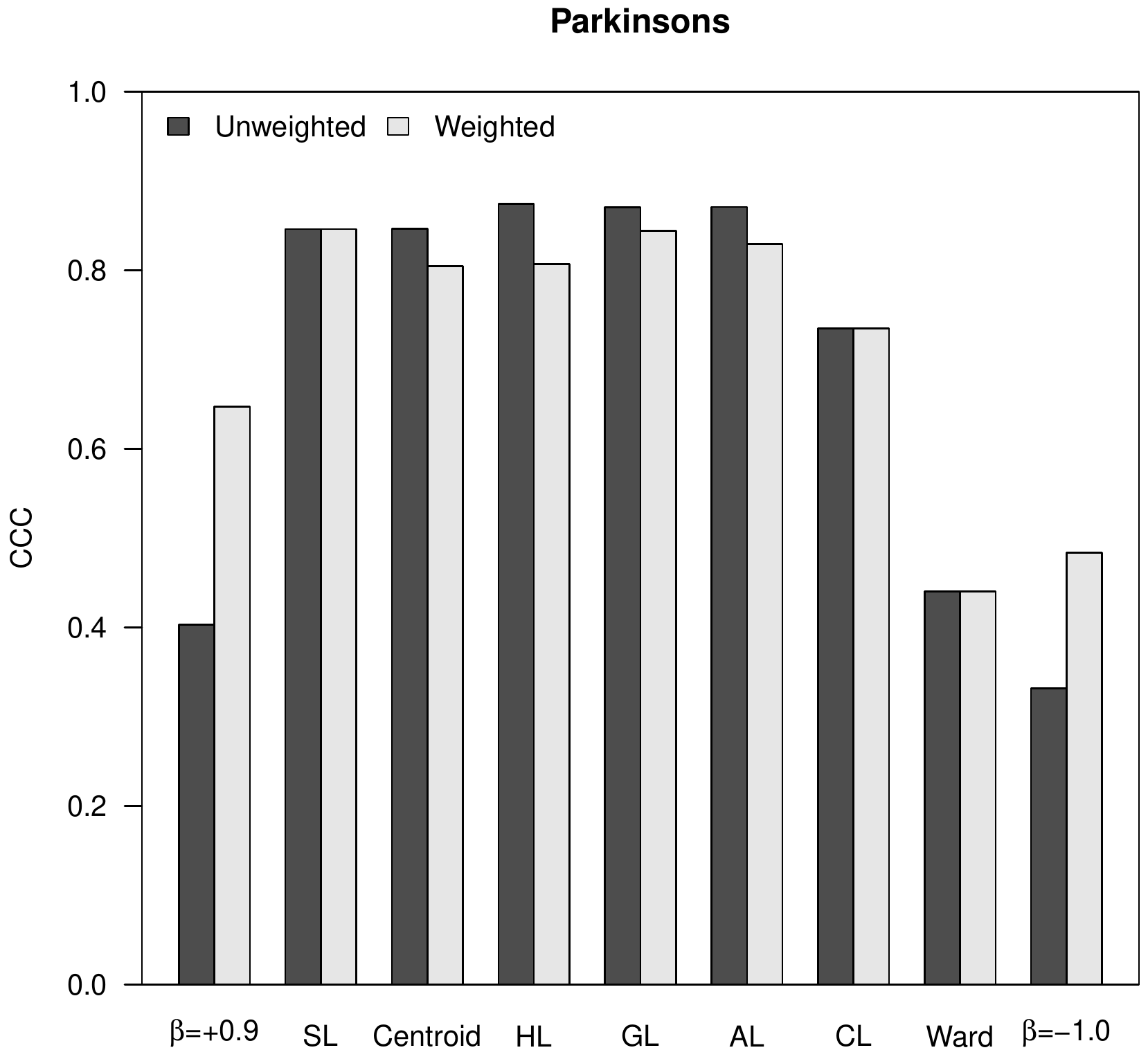}
    \end{tabular}
	\caption{Cophenetic correlation coefficient (CCC). Weighted and unweighted versions of the clustering strategies are compared.}
	\label{fig:cophenetic}
  \end{center}
\end{figure}

In the analysis shown in Figure~\ref{fig:cophenetic}, the CCC is higher for Pythagorean linkages (i.e., HL, GL and AL), and also the unweighted clustering strategies generally perform better than the weighted ones, corroborating the empirical observation already stated by Sneath and Sokal \cite{Sneath1973}. In the case of the almost flat hierarchical trees obtained with $\beta$-flexible clustering when $\beta = +1$, the CCC is very close to $0$.

\subsection{Mean Absolute Error}

\begin{figure}[tb!]
  \begin{center}
    \begin{tabular}{cc}
      \includegraphics[width=0.45\textwidth]{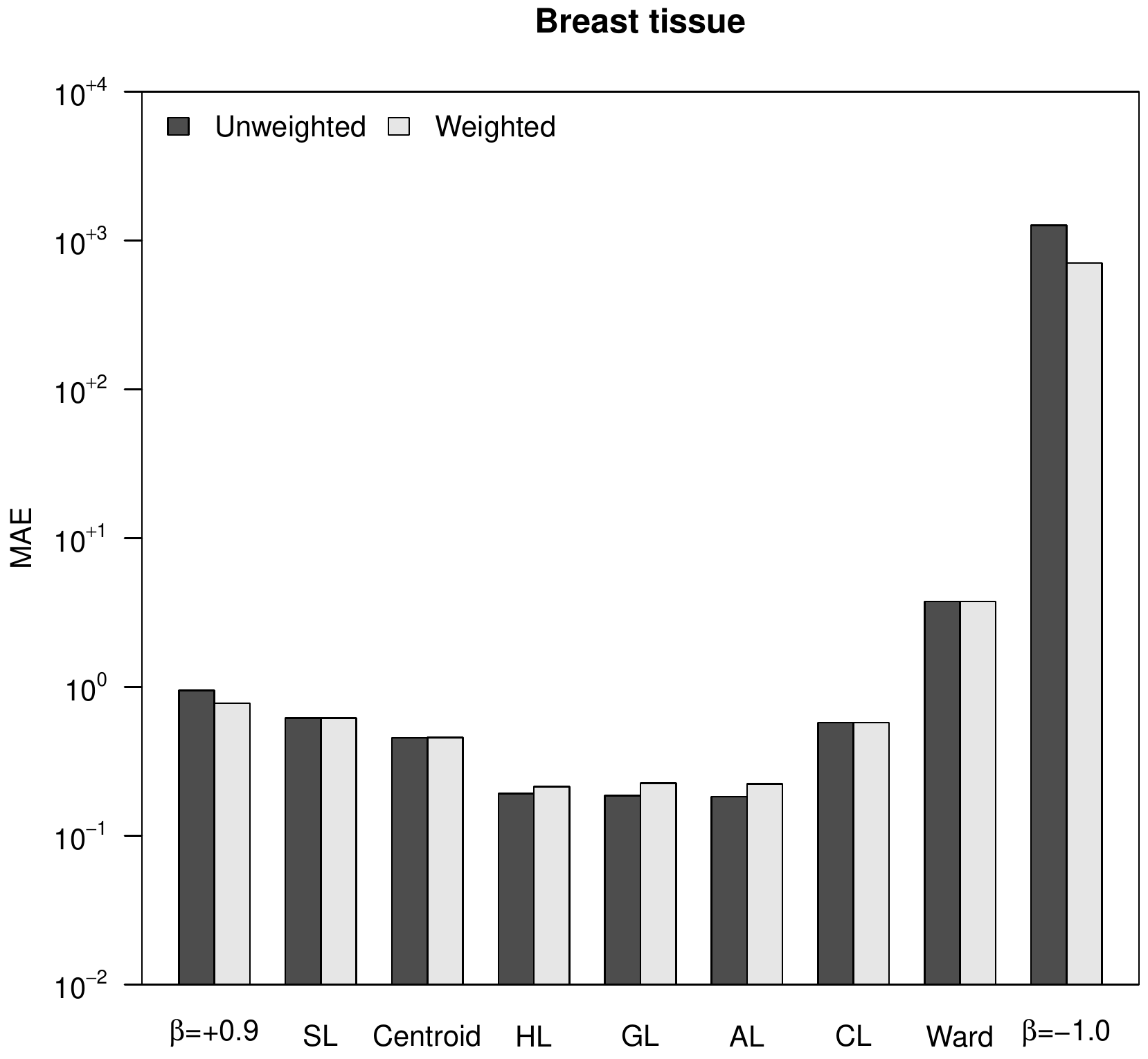} &
      \includegraphics[width=0.45\textwidth]{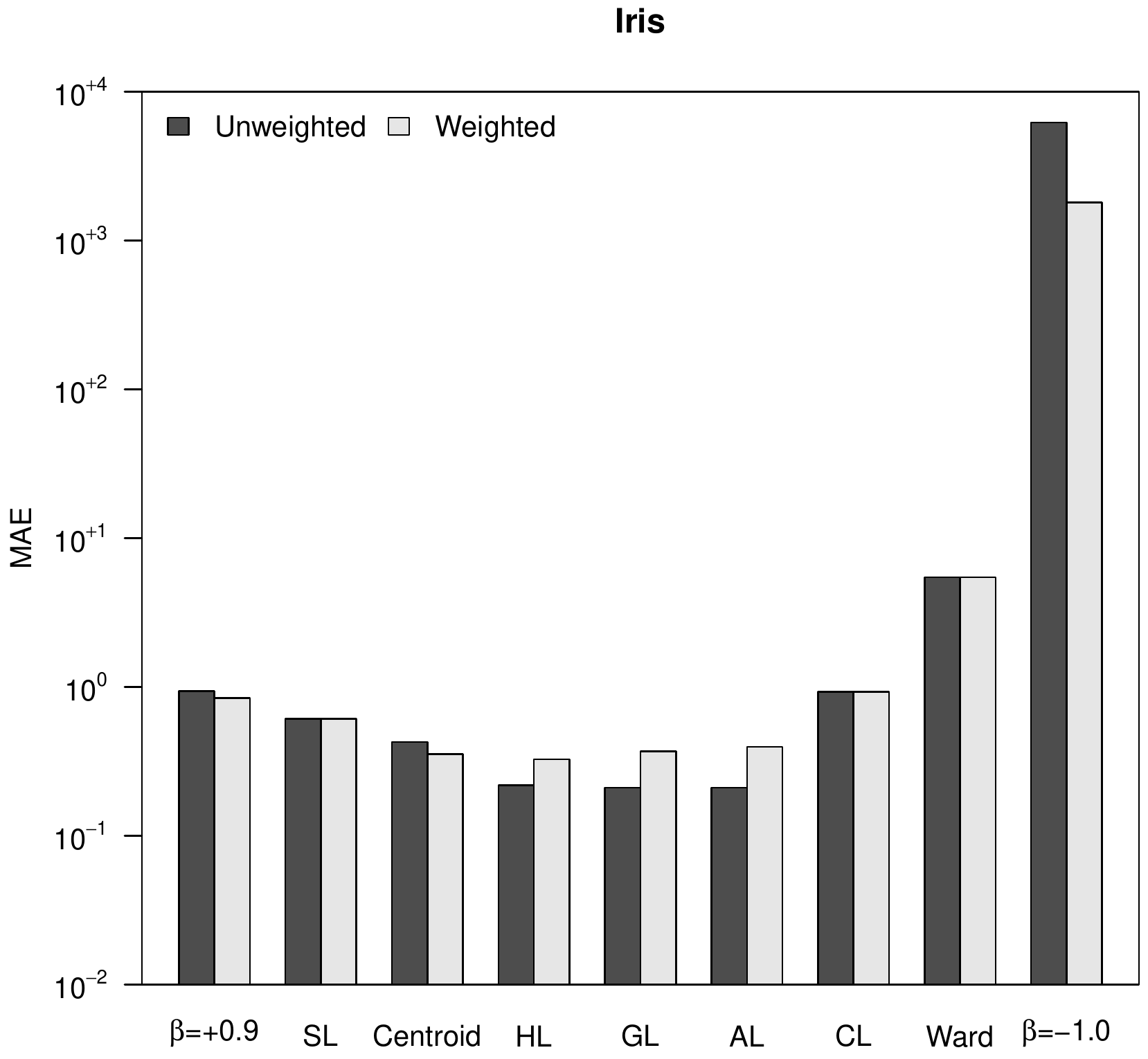} \\
      \includegraphics[width=0.45\textwidth]{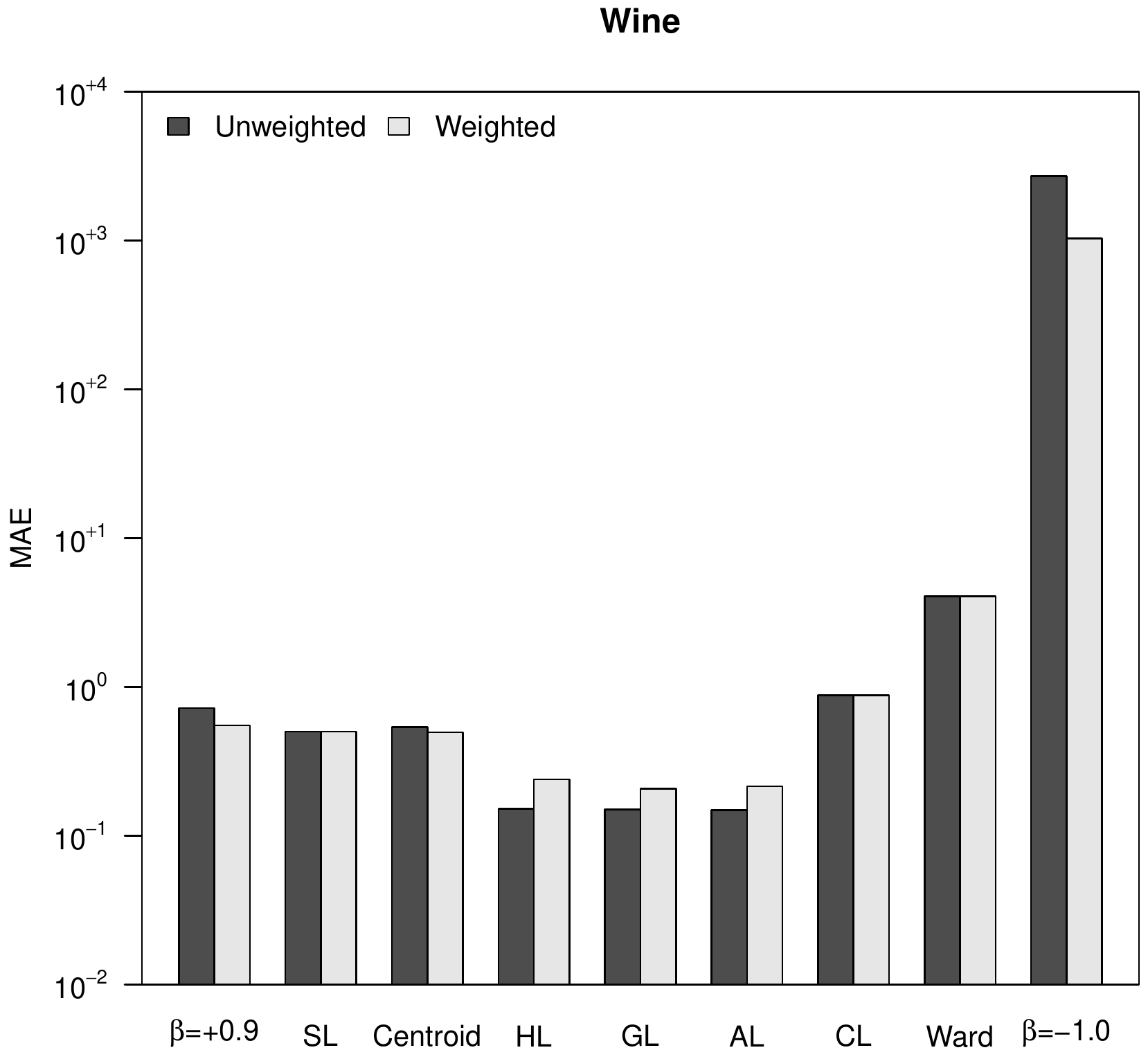} &
      \includegraphics[width=0.45\textwidth]{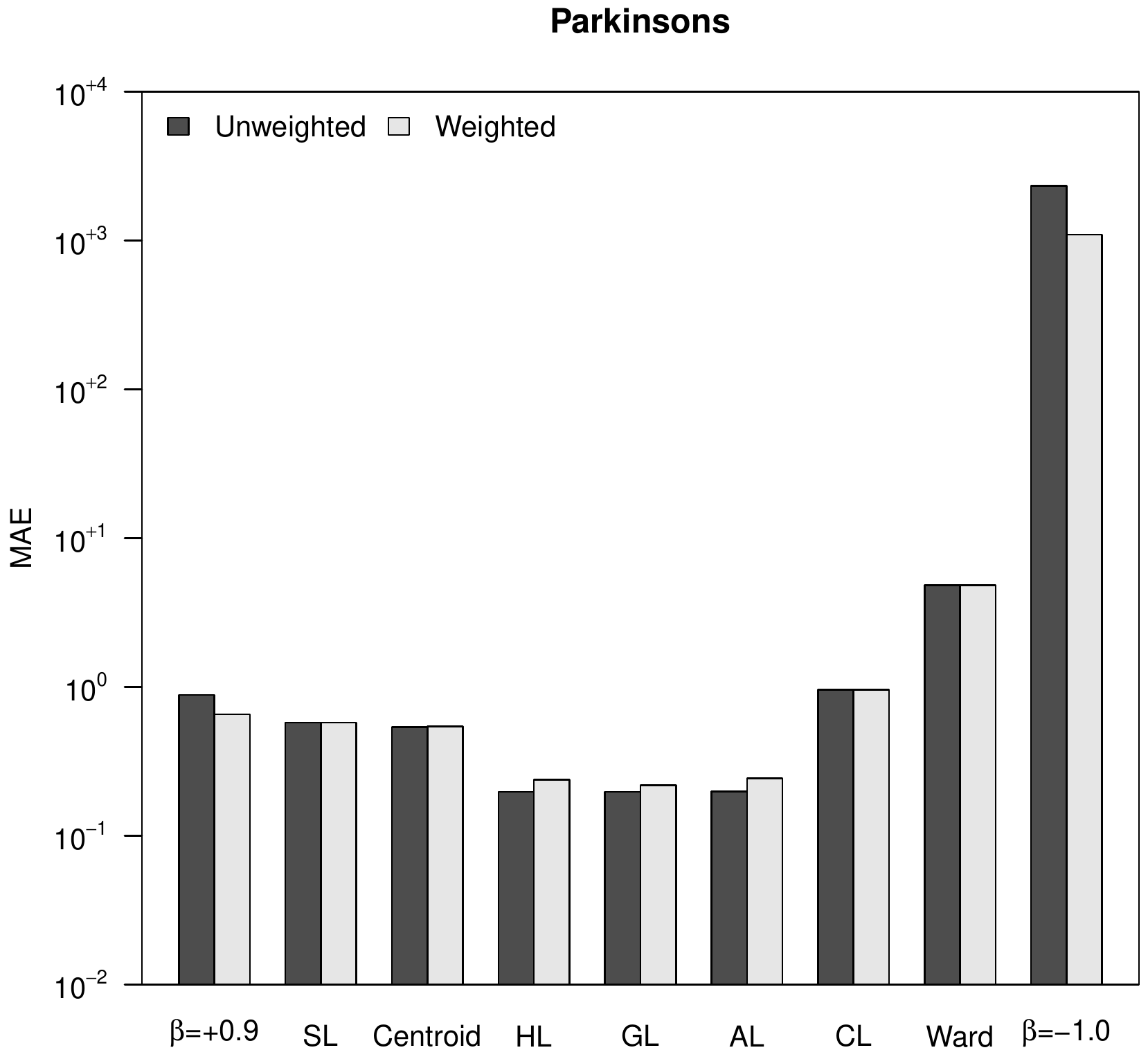}
    \end{tabular}
	\caption{Normalized mean absolute error (MAE), in logarithmic scale. Weighted and unweighted versions of the clustering strategies are compared.}
	\label{fig:mae}
  \end{center}
\end{figure}

The CCC is a bounded measure that does not take into account how different the magnitudes of the distances in the initial matrix are from the distances in the final ultrametric matrix. For this reason, in Figure~\ref{fig:mae} we show the normalized mean absolute error (MAE), which takes into account this type of differences. Note that in the case of the Iris dataset, Ward's method and $\beta$-flexible clustering with $\beta = -1$ showed a very good CCC in Figure~\ref{fig:cophenetic}, while their MAE observed in Figure~\ref{fig:mae} are the worst ones. As a matter of fact, $\beta$-flexible clustering with $\beta = -1$ yields results orders of magnitude worse than all the other methods, for the four datasets shown in Figure~\ref{fig:mae}. The best results are obtained again with Pythagorean linkages, and also unweighted clustering strategies are slightly better than the weighted ones.

\subsection{Space Distortion}

For any agglomerative hierarchical clustering strategy, the initial distances between individuals may be regarded as defining a space with known properties \cite{Lance1967}. When clusters begin to form, if the new distances between clusters are kept within the limits of the same space, then the original model remains unchanged and the clustering strategy is referred to as space-conserving. Otherwise, the clustering strategy is referred to as space-distorting. According to the formalization of the concept of space distortion \cite{Dubien1979}, a clustering strategy is said to be space-conserving if
\begin{equation}
  \label{eq:space_conserving}
  \min_{i \in I} \, \min_{j \in J} \, D(X_{i},X_{j}) \leqslant D(X_{I},X_{J}) \leqslant \max_{i \in I} \, \max_{j \in J} \, D(X_{i},X_{j})\,.
\end{equation}
On the contrary, a clustering strategy is space-contracting if the left inequality, delimited by SL, is not satisfied; and a clustering strategy is space-dilating if the right inequality, delimited by CL, is not satisfied. For space-contracting clustering strategies, as clusters grow in size, they move closer to other clusters. This effect is called chaining and it refers to the successive addition of elements to an ever expanding single cluster \cite{Lance1967}. Space-dilating clustering strategies produce the opposite effect, i.e., clusters moving further away from other clusters as they grow in size.

\begin{figure}[tb!]
  \begin{center}
    \begin{tabular}{cc}
      \includegraphics[width=0.45\textwidth]{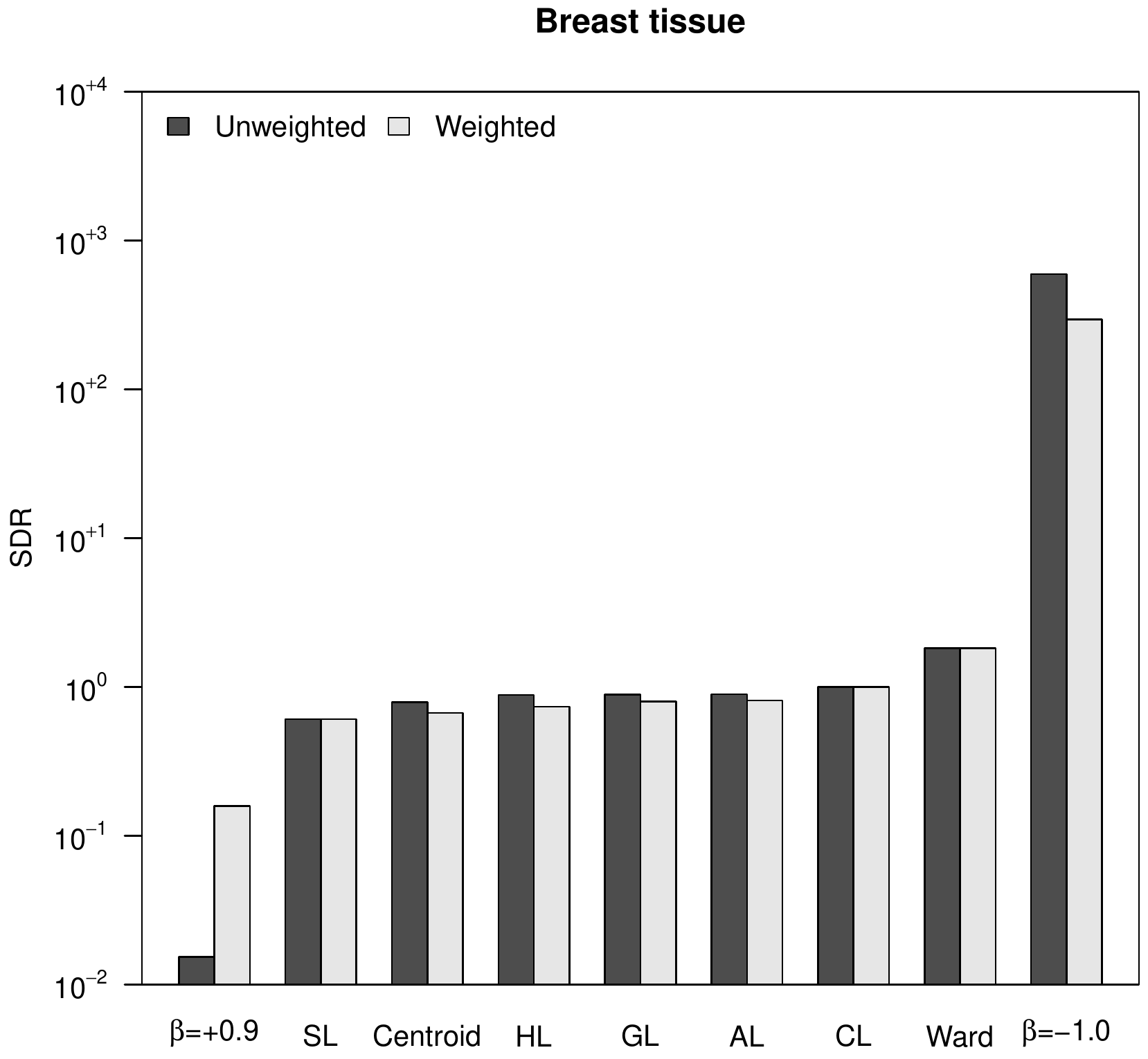} &
      \includegraphics[width=0.45\textwidth]{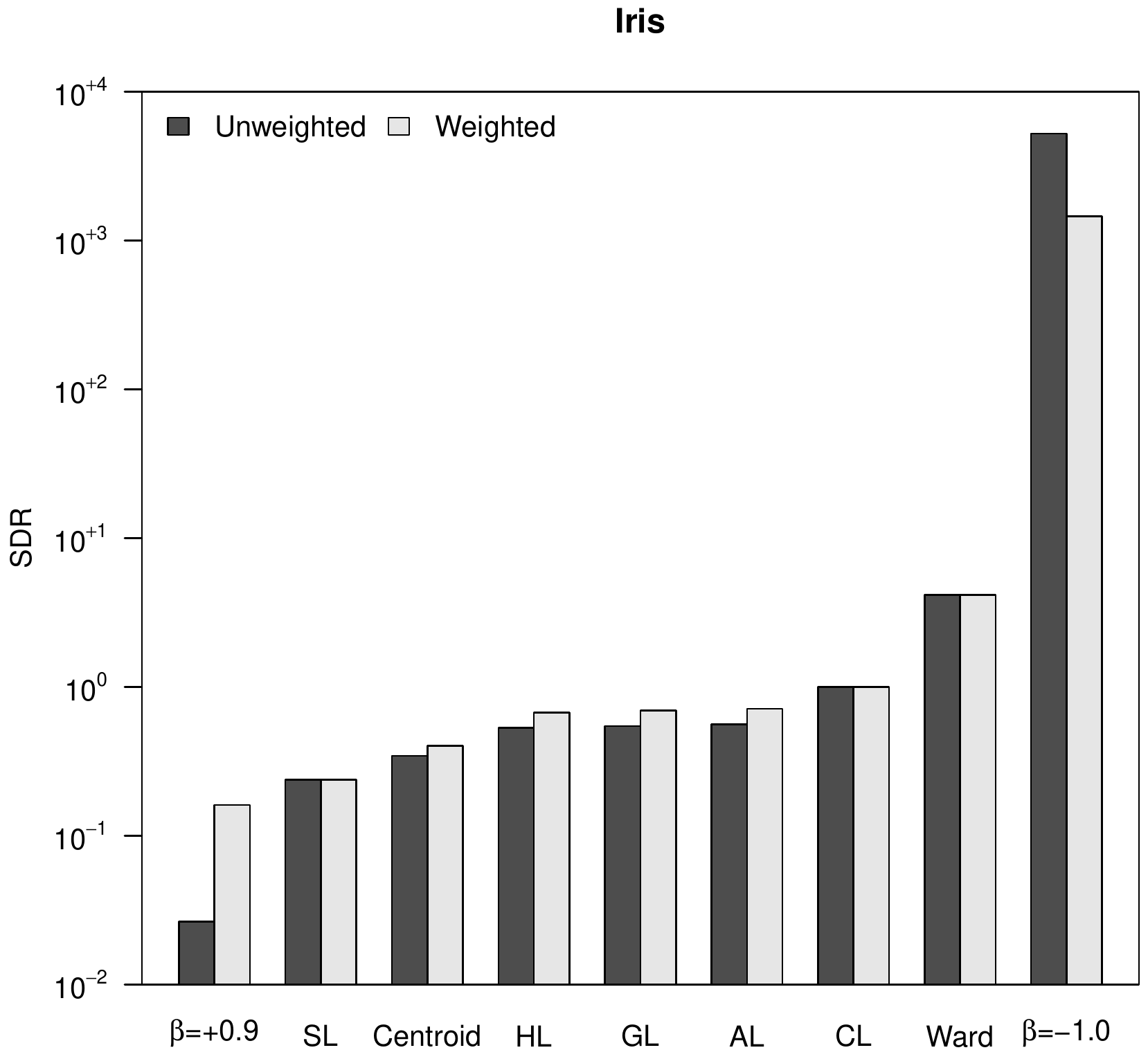} \\
      \includegraphics[width=0.45\textwidth]{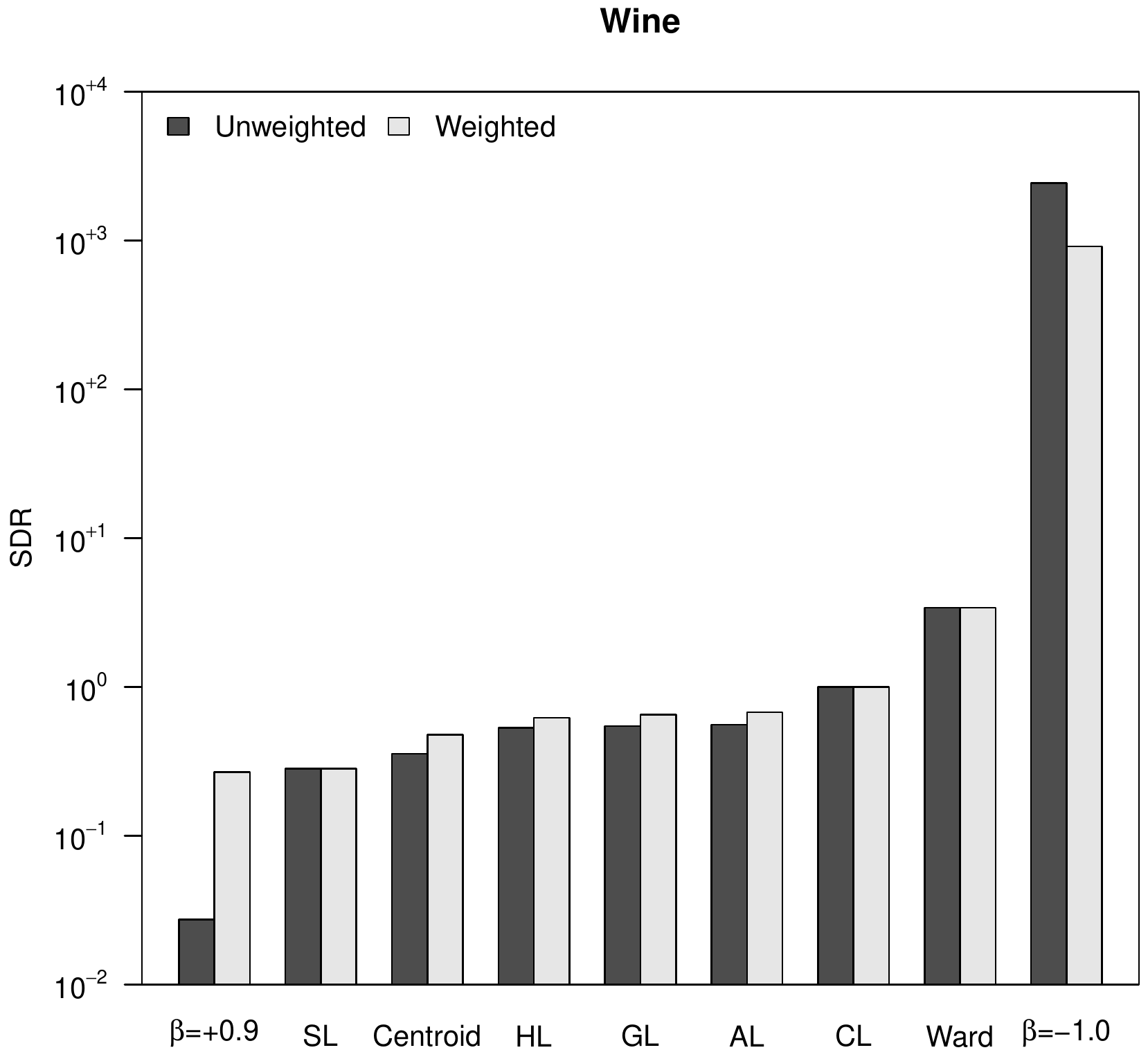} &
      \includegraphics[width=0.45\textwidth]{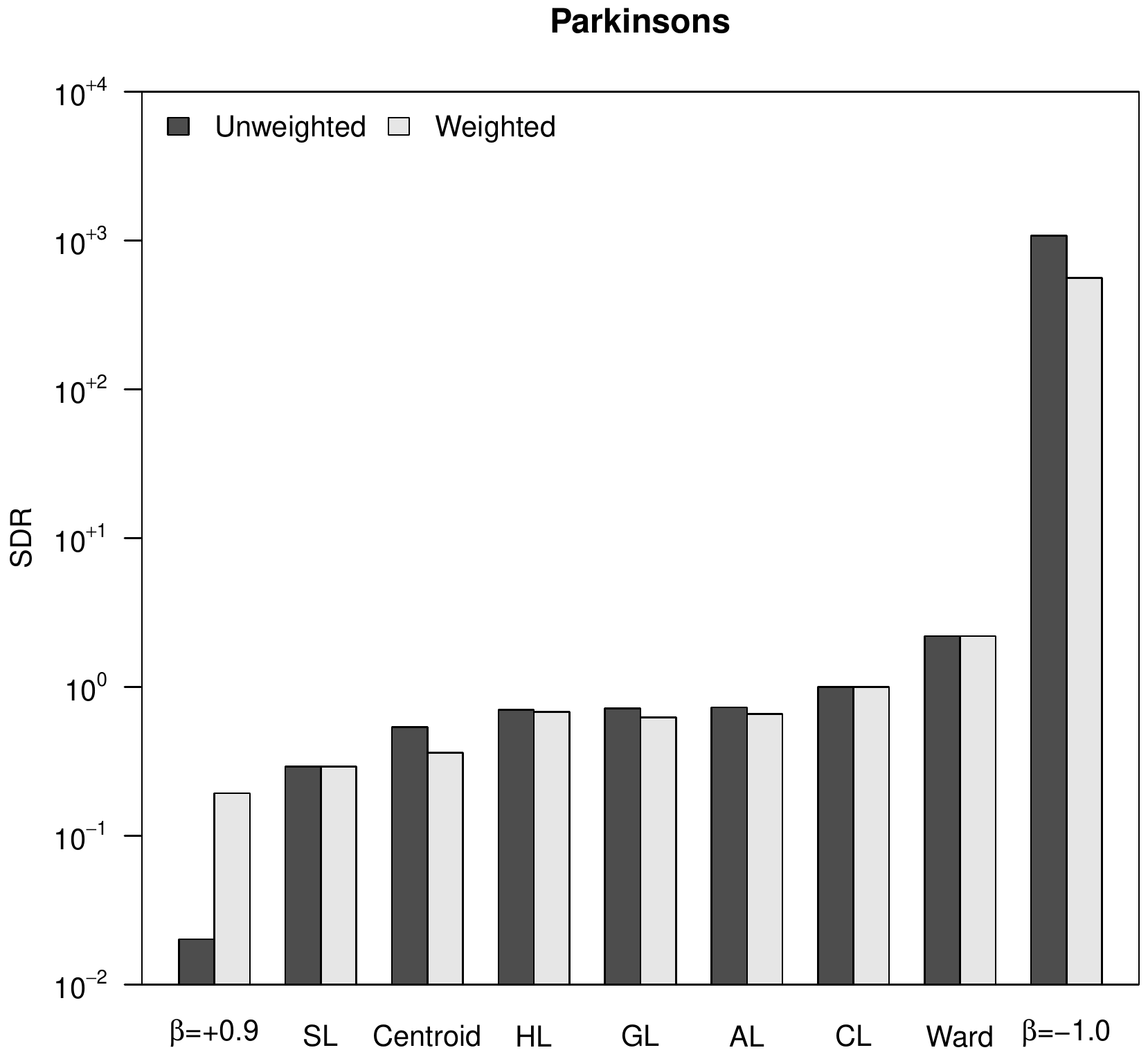}
    \end{tabular}
	\caption{Space distortion ratio (SDR), in logarithmic scale. Weighted and unweighted versions of the clustering strategies are compared.}
	\label{fig:distortion}
  \end{center}
\end{figure}

To numerically assess space distortion, we propose a \textit{space distortion ratio} (SDR) measure, calculated as the quotient between the range of final ultrametric distances, $u(x_{i},x_{j})$, and the range of initial distances, $d(x_{i},x_{j})$:
\begin{equation}
  \label{eq:distortion}
  \mbox{SDR}(u,d) = \frac{\max u(x_{i},x_{j})-\min u(x_{i},x_{j})}{\max d(x_{i},x_{j})-\min d(x_{i},x_{j})}\,.
\end{equation}
The SDR is equal to $1$ for CL, thus this value separates space-conserving hierarchical trees from space-dilating ones. Figure~\ref{fig:distortion} shows the SDR values corresponding to our four case studies. The outstanding differences between initial distances and ultrametric distances in the case of Ward's method and $\beta$-flexible clustering with $\beta = -1$, already observed in Figure~\ref{fig:mae}, allow the classification of both hierarchical clustering methods as space-dilating. With regard to weighting, it cannot be stated that neither weighted nor unweighted clustering strategies produce more space distortion: it depends on the particular dataset.

In Figure~\ref{fig:distortion} it can also be observed the increasing space distortion when $\beta$ decreases in $\beta$-flexible clustering, or when the power $p$ increases in versatile linkage clustering. Both parameters, $\beta$ and $p$, work as cluster intensity coefficients in their respective clustering systems. In the case of versatile linkage, the increasing space distortion when the power $p$ increases is explained by the generalized mean inequality in Equation~\ref{eq:generalized_inequality}. Therefore, taking also into account that, according to Equation~\ref{eq:space_conserving}, space-conserving clustering strategies are lower bounded by SL ($p \rightarrow -\infty$) and upper bounded by CL ($p \rightarrow +\infty$), we can state that versatile linkage defines an infinite system of space-conserving strategies for agglomerative hierarchical clustering.

\subsection{Tree Balance}

\begin{figure}[tb!]
  \begin{center}
    \begin{tabular}{cc}
      \includegraphics[width=0.45\textwidth]{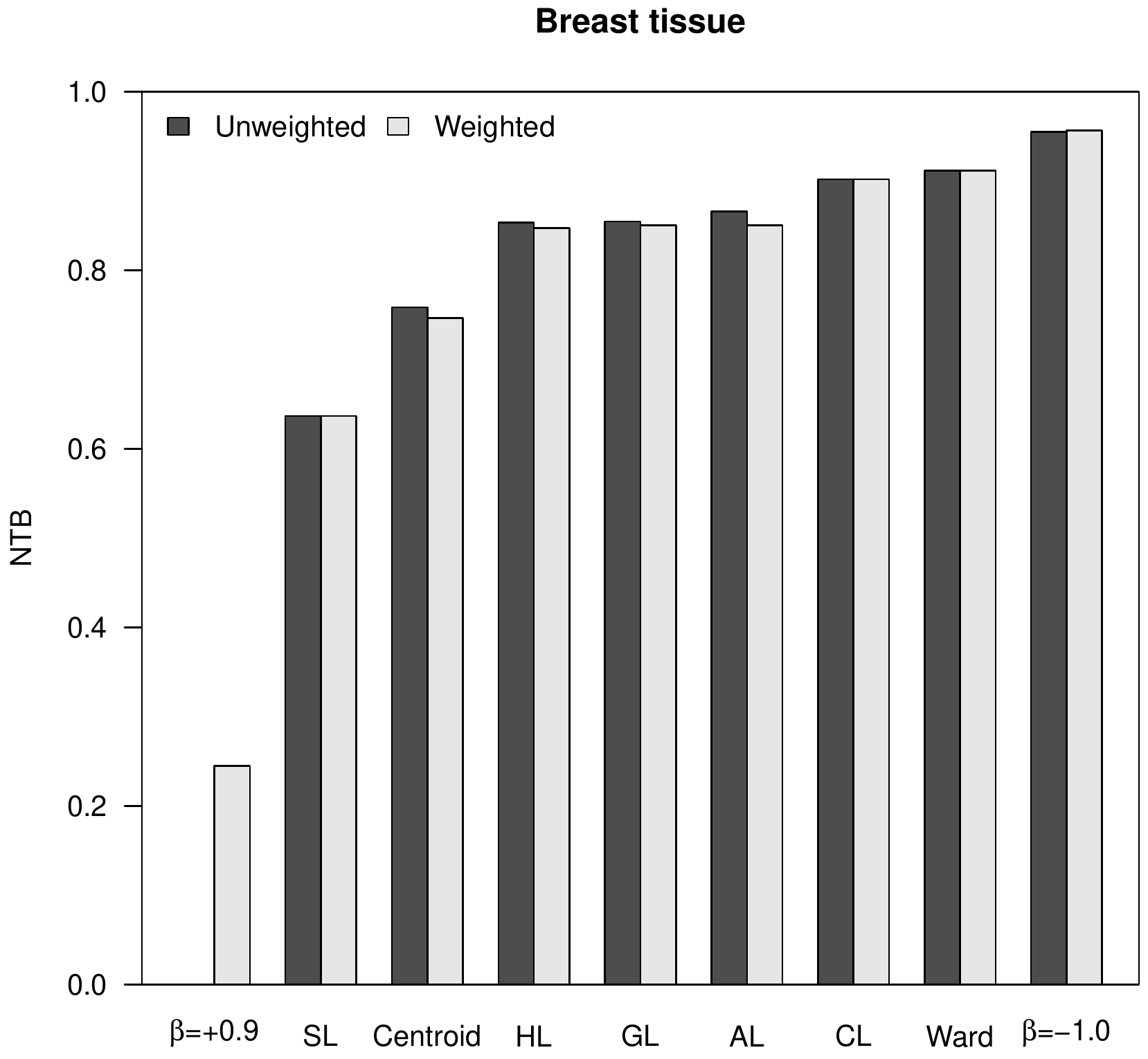} &
      \includegraphics[width=0.45\textwidth]{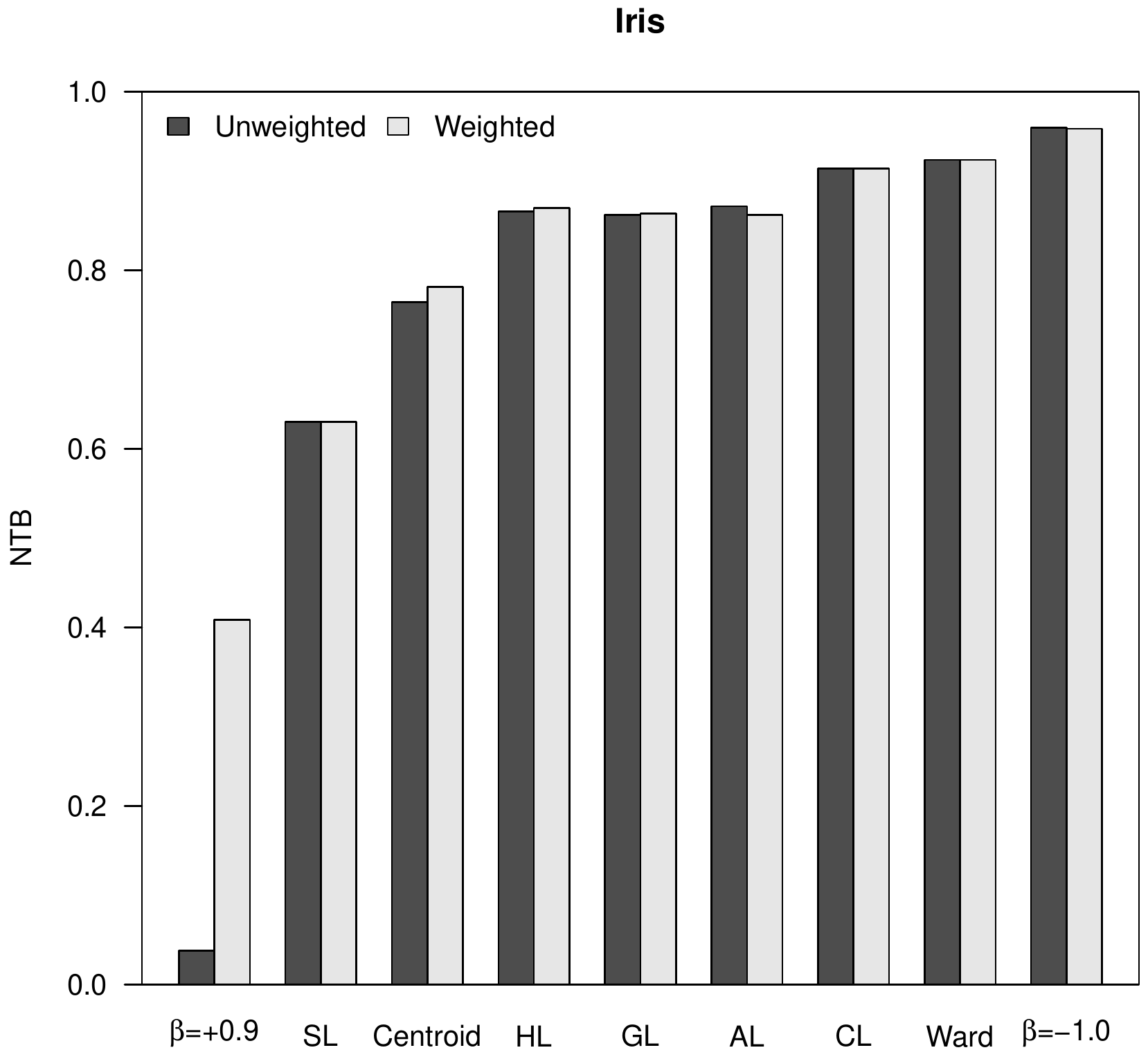} \\
      \includegraphics[width=0.45\textwidth]{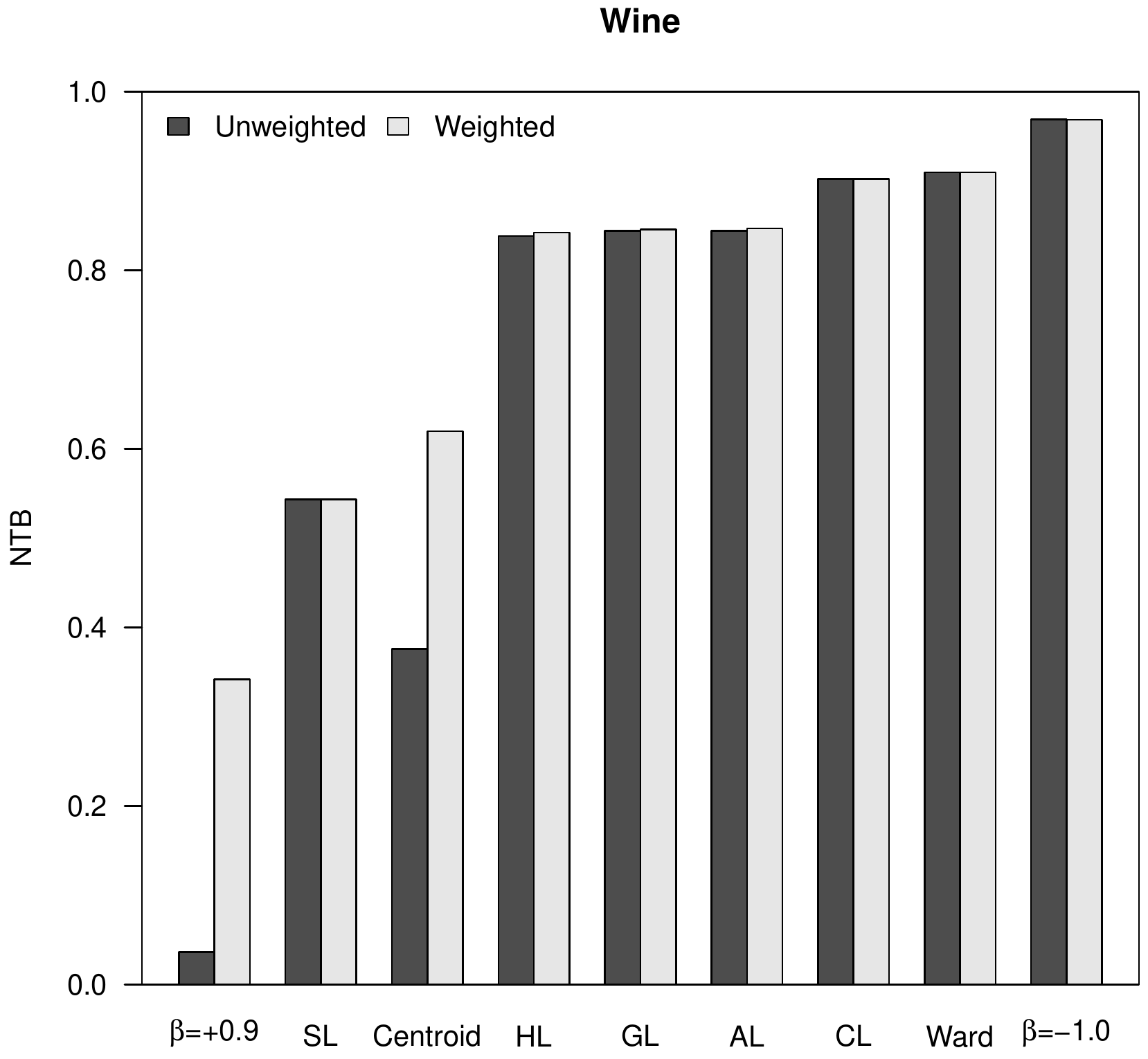} &
      \includegraphics[width=0.45\textwidth]{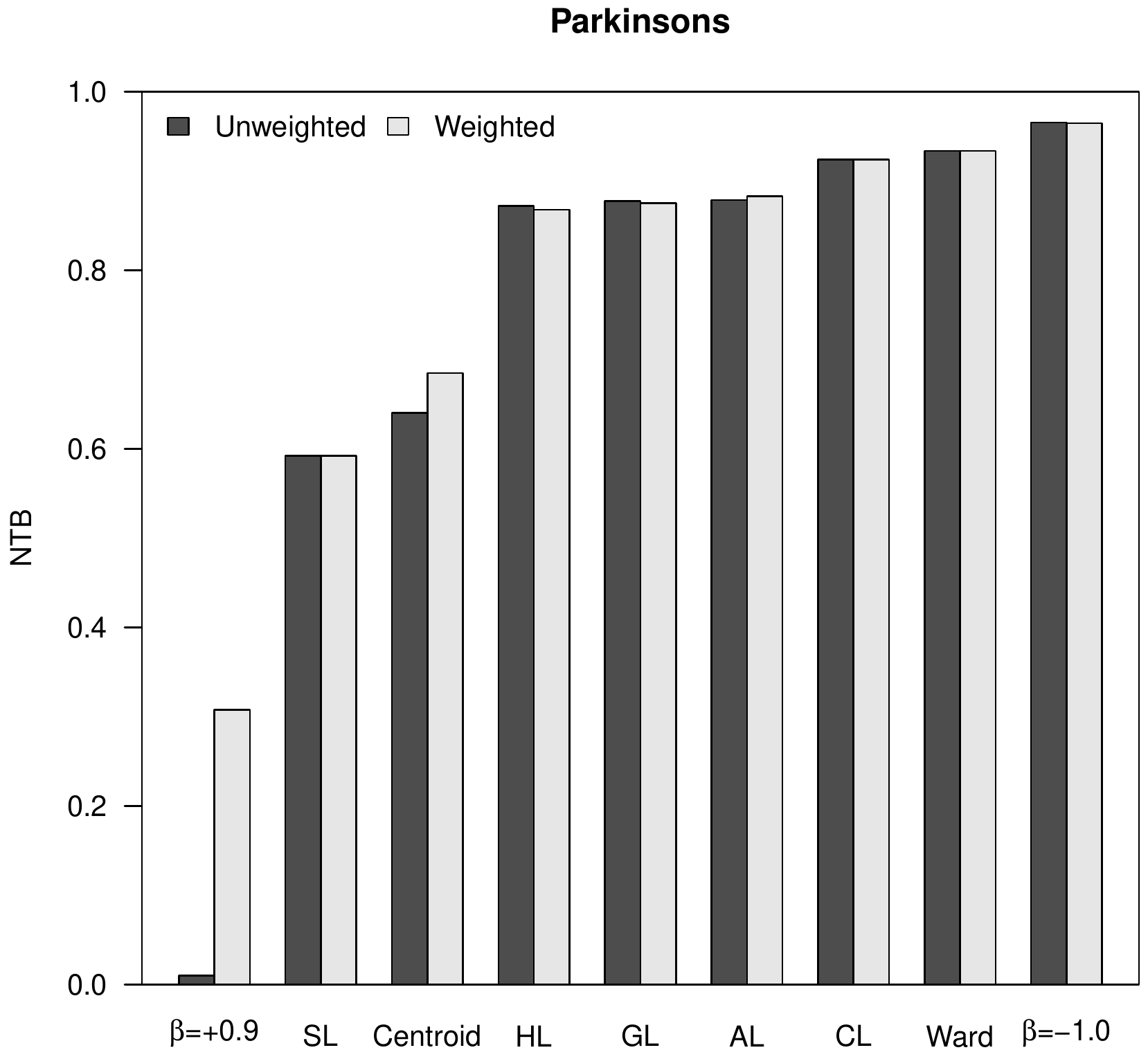}
    \end{tabular}
	\caption{Normalized tree balance (NTB). Weighted and unweighted versions of the clustering strategies are compared.}
	\label{fig:balance}
  \end{center}
\end{figure}

We use the concept of entropy from information theory, more concretely Shannon's entropy \cite{Shannon1948}, to introduce a new measure to assess the degree of homogeneity in size of the clusters in a hierarchical tree. Given a cluster $X_{I}$, we define its entropy as
\begin{equation}
  H_{I} = - \sum_{i \in I} p_{i} \log_{|I|} (p_{i})\,,
\end{equation}
where $p_{i} = \frac{|X_{i}|}{|X_{I}|}$ is the proportion of individuals in cluster $X_{I}$ that are also members of subcluster $X_{i}$. Next, we define the \textit{tree balance}, $H$, of a hierarchical tree as the average entropy of all its internal clusters. The maximum tree balance is equal to $1$ and it is obtained, for instance, for a completely flat hierarchical tree with a single cluster containing the $N$ individuals in the collection. Another example of hierarchical trees with maximum tree balance are the regular $m$-way trees obtained when applying the Baire-based divisive hierarchical clustering algorithm on a collection of sequences with uniformly distributed prefixes \cite{Bradley2010, Contreras2012}. On the contrary, the minimum tree balance, $H_{\min}$, corresponds to a binary tree where individuals are chained one at a time:
\begin{equation}
  H_{\min} = \frac{1}{N - 1} \left[ \log_{2}(N) + \sum_{n = 2}^{N - 1} \frac{1}{n + 1}\log_{2}(n) \right]\,.
\end{equation}
Now, we can define the \textit{normalized tree balance} (NTB) as
\begin{equation}
  \mbox{NTB} = \frac{H - H_{\min}}{1 - H_{\min}}\,,
\end{equation}
which becomes a measure with values between $0$ and $1$. Figure~\ref{fig:balance} shows the NTB values obtained for our case studies. Similarly to space distortion, tree balance increases when $\beta$ decreases in $\beta$-flexible clustering, or when the power $p$ increases in versatile linkage clustering. In the case of the almost flat hierarchical trees obtained with $\beta$-flexible clustering when $\beta = +1$, the NTB is very close to $1$. Finally, according to the values observed in Figure~\ref{fig:balance}, it cannot be stated that neither weighted nor unweighted clustering strategies produce hierarchical trees that are more balanced.

\section{Conclusions}
\label{sec:conclusions}

Agglomerative hierarchical clustering methods have been continually evolving since their origins back in the 1950s, and historically they have been deployed in very diverse application domains, such as geosciences, biosciences, ecology, chemistry, text mining and information retrieval, among others \cite{Murtagh2017b}. Nowadays, with the advent of the big data revolution, hierarchical clustering methods have had to address the new challenges brought by more recent application domains that require the hierarchical clustering of thousands of observations \cite{Murtagh2017a}.

In this work we have introduced versatile linkage, an infinite family of agglomerative hierarchical clustering strategies based on the definition of generalized mean. We have shown that the versatile linkage family contains as particular cases not only the traditionally important strategies of single linkage, complete linkage and arithmetic linkage, but also two new clustering strategies such as geometric linkage and harmonic linkage. In addition, we have given both weighted and unweighted versions of these hierarchical clustering strategies, and we have proved the monotonicity of versatile linkage strategies, which guarantees the absence of inversions in the hierarchy. Although we have built versatile linkage upon the multidendrograms variable-group methods to ensure the uniqueness of the clustering, it may also be used with the common pair-group approach just by breaking ties randomly.

We have shown that any descriptive analysis of hierarchical trees in terms of cophenetic correlation should be complemented with the use of other measures capable of describing the space distortion that different hierarchical clustering strategies cause. Under this point of view, we have shown that it is helpful to use other measures such as the mean absolute error or the space distortion ratio. The latter, in addition, provides a way to describe numerically the increase in space distortion observed all along a system of hierarchical clustering strategies such as versatile linkage.

Space distortion is inversely proportional to clustering intensity: space-contracting clustering strategies drive systems to cluster very weakly and produce a chaining effect, while space-dilating clustering strategies drive systems to cluster with high intensity and produce very compact clusters. These differences are described by the normalized tree balance measure introduced here, which is based on Shannon's entropy. Tree balance and space distortion are two new descriptive measures meant to be helpful to analyze and understand any hierarchical tree.

The $\beta$-flexible clustering also integrates an infinite number of agglomerative hierarchical clustering strategies into a single system, driven by a parameter $\beta$ that works as a cluster intensity coefficient. However, to the best of our knowledge, no one has rigorously defined yet a range of values of $\beta$ for which the corresponding $\beta$-flexible clustering strategies can be regarded as space-conserving. Unlike the $\beta$-flexible clustering system, we have shown that the versatile linkage family is space-conserving.

\section*{Acknowledgements}

This work has been partially supported by MINECO through grant FIS2015-71582-C2-1 (S.G.), Generalitat de Catalunya project 2017-SGR-896 (S.G.), and Universitat Rovira i Virgili projects 2017PFR-URV-B2-29 (A.F.) and 2017PFR-URV-B2-41 (S.G.).


\end{document}